\documentclass{aa}
\usepackage[varg]{txfonts}
\usepackage{color}
\usepackage{graphicx}
\usepackage{natbib}
\usepackage{epstopdf}
\bibpunct{(}{)}{;}{a}{}{,} %% natbib format like A&A and ApJ

\begin{document}

\title{Damping of slow magnetoacoustic oscillations by the misbalance between heating and cooling processes in the solar corona}
\author{\ D.~Y.~Kolotkov\inst{1}\thanks{Corresponding~author: D.~Y.~Kolotkov,~D.Kolotkov.1@warwick.ac.uk} \and V.~M.~Nakariakov\inst{1,2}\and D.~I.~Zavershinskii\inst{3,4}}

\authorrunning{Kolotkov et al.}
\titlerunning{Damping of slow waves by the thermal misbalance}

\institute{Centre for Fusion, Space and Astrophysics, Department of Physics, University of Warwick, CV4 7AL, UK\label{1}
\and
St. Petersburg Branch, Special Astrophysical Observatory, Russian Academy of Sciences, 196140, St. Petersburg, Russia \label{2}
\and
Samara National Research University, Department of Physics, Samara 443086, Russia\label{3}
\and
Lebedev Physical Institute of Russian Academy of Sciences, Samara Branch, Department of Theoretical physics\label{4}
}

\date{Received \today /Accepted dd mm yyyy}

\abstract
{Rapidly decaying slow magnetoacoustic waves are regularly observed in the solar coronal structures, offering a promising tool for a seismological diagnostics of the coronal plasma, including its thermodynamical properties. }
{The effect of damping of standing slow magnetoacoustic oscillations in the solar coronal loops is investigated accounting for the field-aligned thermal conductivity and a wave-induced misbalance between radiative cooling and some unspecified heating rates.}
{The non-adiabatic terms were allowed to be arbitrarily large, corresponding to the observed values. The thermal conductivity was taken in its classical form, and a power-law dependence of the heating function on the density and temperature was assumed. {The analysis was conducted in the linear regime and in the infinite magnetic field approximation.}}
{The wave dynamics is found to be highly sensitive to the characteristic time scales of the thermal misbalance. Depending on certain values of the misbalance time scales three regimes of the wave evolution were identified, namely the regime of a suppressed damping, enhanced damping where the damping rate drops down to the observational values, and acoustic over-stability. The specific regime is determined by the dependences of the radiative cooling and heating functions on thermodynamical parameters of the plasma in the vicinity of the perturbed thermal equilibrium.}
{The comparison of the observed and theoretically derived decay times and oscillation periods allows us to constrain the coronal heating function. For typical coronal parameters, the observed properties of standing slow magnetoacoustic oscillations could be readily reproduced with a reasonable choice of the heating function.}


\maketitle

\section{Introduction}

The study of wave and oscillatory processes in the plasma of the solar corona is one of the most rapidly developing research topics of modern solar physics \citep[e.g.][]{2012RSPTA.370.3193D, 2016GMS...216..395W}. The interest in coronal oscillations is connected, in particular, with their seismological potential, i.e. with the use of the oscillations as natural probes of the plasma and physical processes operating there \citep[e.g.][]{2014SoPh..289.3233L}. Moreover, the striking similarity between the properties of oscillations detected in solar and stellar flares \citep[see, e.g.][]{2016ApJ...830..110C}, suggests interesting perspectives for the exploitation of the solar-stellar analogy.

Slow magnetoacoustic waves are often detected in coronal plasma non-uniformities, such as  coronal loops, and plumes and the interplume regions, as propagating periodic disturbances of the EUV emission, \citep[see, e.g.][respectively]{2009SSRv..149...65D, 2016GMS...216..419B}. Another common manifestation of slow waves in the corona are standing waves in loops, detected as rapidly decaying periodic Doppler shifts of coronal emission lines \citep[see, e.g.][]{2011SSRv..158..397W}. Standing slow waves are usually refereed to as SUMER oscillations, after the instrument used in their first detection \citep[SoHO/SUMER, see][]{2002ApJ...574L.101W} and interpretation \citep{2002ApJ...580L..85O}. SUMER oscillations still remain a subject of intensive studies. For example, standing slow waves in non-flaring fan loops, with the periods of 27~min, damping time about 45~min, and the phase speed corresponding to the plasma temperature of about 0.6~MK, have been studied by \cite{2017ApJ...847L...5P}. A 10-min periodicity has been identified in the time series of Doppler shift and line-integrated intensity of the \ion{Fe}{xxi} emission line, soft X-ray flux, and EUV light curves \citep{2017ApJ...849..113L}. A 2-min oscillation of the thermal component of the microwave emission of a solar flare has been interpreted in terms of the emission modulation by a standing slow wave. An 80~s oscillation of  the X-ray and microwave emissions in a solar flare has been associated with second harmonic of standing slow wave in a flaring arcade \citep{2019MNRAS.483.5499K}. 
The 8--30~min periodic pulsations of the soft X-ray emission generated in an active region before a flare could also be associated with standing slow waves \citep{2016ApJ...833..206T}.
Seismological applications of slow waves include the estimation of the polytropic index \citep{2011ApJ...727L..32V, 2018ApJ...868..149K}, average magnetic field \citep{2007ApJ...656..598W} in the oscillating loop,  and transport coefficients \citep{2015ApJ...811L..13W, 2018ApJ...860..107W}. An important foundation of the interpretations and seismology is provided by the forward modelling of imaging and spectroscopic observables \citep{2015ApJ...807...98Y}.

Recent theoretical studies of standing slow waves in coronal loops include accounting for weakly-nonlinear effects that are found to manifest as an appearance of higher parallel harmonics \citep[e.g.][]{2016ApJ...824....8K}; full-MHD numerical simulations with various scenario of transport processes, which aim at revealing the reason for the unexpected linear scaling of the observed damping time with the oscillation period \citep[e.g.][]{2018ApJ...860..107W},  and the excitation mechanism \citep[e.g.][]{2018AdSpR..61..645P}.

%Theoretical modelling of standing slow waves is usually performed in the infinite magnetic field limit of the thin flux tube approximation, i.e., the set of the governing equations coincides with the one-dimensional acoustics.

An important physical process that should be taken into account in the modelling of compressive oscillations is the perturbation of the thermal equilibrium by the oscillation, i.e. the effect of the misbalance between radiative and, possibly, thermal conductive cooling, and an unspecified but definitely present heating. Similar effects are considered in the interstellar medium and molecular clouds, while mainly in the contexts of the plasma condensation caused by thermal instability \citep[e.g.][]{2017MNRAS.469.1403K}, and basic theoretical studies of the autowave regimes \citep[e.g.][]{2013TePhL..39..676Z} and  Alfv\'en wave amplification \citep[e.g.][]{2014TePhL..40..701Z}.
In the coronal context, it has been shown that the effect of thermal misbalance can either strengthen the damping or suppress it \citep[e.g.][]{2017ApJ...849...62N}. However, this conclusion was reached in the limit of weak non-adiabaticity, using the assumption that the imaginary part of the oscillation frequency is much smaller than the real part. On the other hand, for example, the damping time of SUMER oscillations is known to be comparable with the oscillation period. It justifies the need for softening this assumption. 

{The aim of this paper is to develop a theory of linear standing slow magnetoacoustic oscillations in coronal loops with thermal misbalance.} In Section~\ref{sec:gov} we describe the model, and derive dispersion relations that are analysed in Section~\ref{sec:stab}. The findings are summarised and discussed in Section~\ref{sec:sumcon}.

\section{Governing equations, time scales, and dispersion relation}
\label{sec:gov}
We consider evolution of slow magnetoacoustic waves in the infinite magnetic field approximation, upon which the set of governing equations reduces to the usual hydrodynamic Euler equation, continuity equation, ideal gas state equation, and the energy equation {(see Eqs.~(\ref{eq_mov})--(\ref{eq:energy_full}), respectively).}
{This approximation is extensively used for modelling slow waves in the corona, see e.g. \citet{2000A&A...362.1151N,2002ApJ...580L..85O,2004A&A...415..705D,2008ApJ...685.1286V,2013A&A...553A..23R,2016ApJ...824....8K}.
%, including textbooks \citet{1982soma.book.....P} and \citet{2004prma.book.....G} with its detailed justification.
Under this approximation, the waves are assumed to propagate strictly along the ambient infinitely stiff magnetic field lines, hence do not perturb the field and their speed is independent of it.}

{Accounting for the effects of the optically thin radiation, unspecified heating, and thermal conductivity, the governing equations are}
\begin{align}\label{eq_mov}
&\rho \frac{d V_{z}}{d t} = -\frac{\partial P}{\partial z},\\
&\frac{\partial \rho}{\partial t} +  \frac{\partial }{\partial z} \left(\rho V_{z} \right) = 0,\\
&P=\frac{k_\mathrm{B} T \rho}{m},\\
&C_\mathrm{V}\frac{dT}{dt} - \frac{k_\mathrm{B}T}{m\rho}\frac{d\rho}{dt}=-Q(\rho,T)+\frac{\kappa}{\rho}\frac{\partial^2T}{\partial z^2},\label{eq:energy_full}
\end{align}
where $\rho$, $T$, and $P$ are the density, temperature, and pressure, respectively; $ V_{z}$ is the velocity component along the \textit{z}-axis which coincides with the magnetic field direction, $k_\mathrm{B}$ is Boltzmann constant, $m$ is the mean particle mass, $C_\mathrm{V} = (\gamma - 1)^{-1}k_\mathrm{B}/m$ is the specific heat capacity at constant volume with $\gamma = 5/3$ being the standard adiabatic index, $\kappa$ is the field-aligned thermal conductivity, and the function $Q(\rho,T) = L(\rho,T) - H(\rho,T)$ combines the effects of radiative losses $L(\rho,T)$ and some unspecified heating $H(\rho,T)$. For the energy equation in form (\ref{eq:energy_full}), the heating/cooling function $Q(\rho,T)$ is measured in W kg$^{-1}$.
{For example, {numerous} observational studies demonstrated that the temperature across and along the loop remains almost constant {(see e.g. \citet{2014LRSP...11....4R} for the detailed review of the coronal loop properties, and} \citet{Gupta2019} and references therein {for the most recent results}). Hence, we consider the plasma to be in a uniform isothermal equilibrium.
%As such, the effect of the thermal conductivity on the initial equilibrium is absent, while it is fully determined by the balance between radiative and heating processes, i.e. by the condition
Thus, in the equilibrium $Q(\rho_0,T_0) = 0$, where the index 0 indicates equilibrium quantities. In general, the equilibrium thermal structure of the loop is also determined by thermal conduction at the footpoints. But, as we consider waves in the coronal, almost isothermal part of an active region, this effect is omitted.
{For the slow waves propagating upwards along loops and plumes this omission is naturally justified. For standing slow waves this omission could be justified by the structure of the pressure, density and temperature perturbations along the loop. In contrast with the perturbations of the parallel velocity that have nodes at the footpoints, perturbations of thermodynamical parameters in standing slow waves have anti-nodes at the footpoints \citep[e.g.][]{2016ApJ...826L..20R,2018ApJ...860..107W}. Hence, near the footpoints the derivative of the temperature perturbation in the wave with respect to the field-aligned coordinate could be taken as zero, {thus suppressing the wave damping by the thermal conduction in these regions.} Thus, in our analysis the chromosphere and transition region act only as the solid-wall perfectly reflecting boundaries for slow waves and are not involved in the wave evolution by any other mean \citep[see e.g.][where a similar approach was employed for the coronal slow wave modelling]{2002ApJ...580L..85O,2005A&A...436..701S,2007ApJ...659L.173T}.}
{In other words, our simple reflecting boundary conditions mimic a more realistic model of the transition region and the chromosphere used by e.g. \cite{2004A&A...414L..25N} or \citet{2016ApJ...826L..20R}, in which slow waves are found to naturally reflect at the lower boundary because they hit the transition region.}
We need to stress that in the considered scenario the waves do not contribute to the heating themselves, but perturb the physical parameters of the plasma that may affect the efficiency of the heating.}

\begin{figure}
	\begin{center}
		\includegraphics[width=0.49\linewidth]{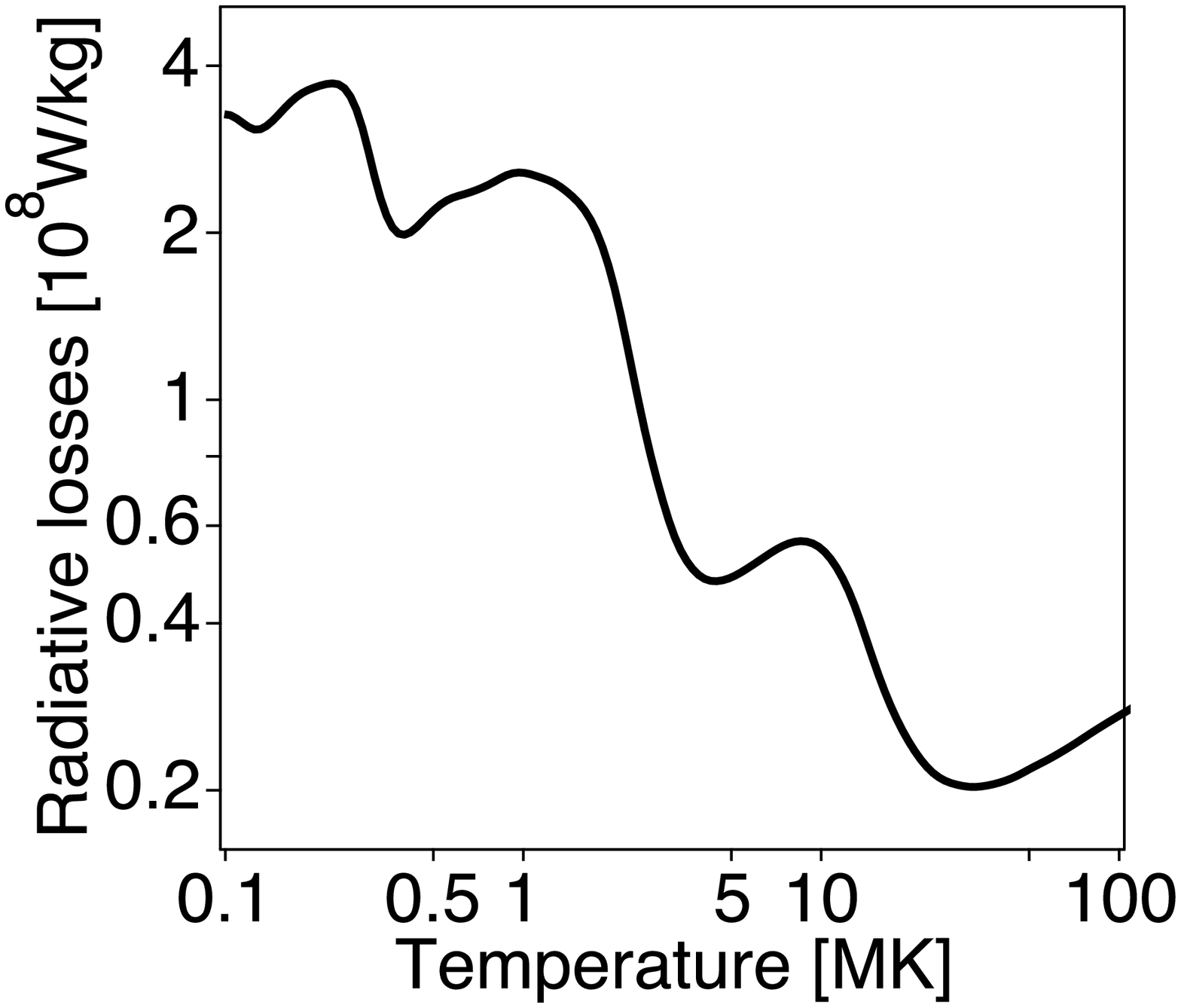}
		\includegraphics[width=0.49\linewidth]{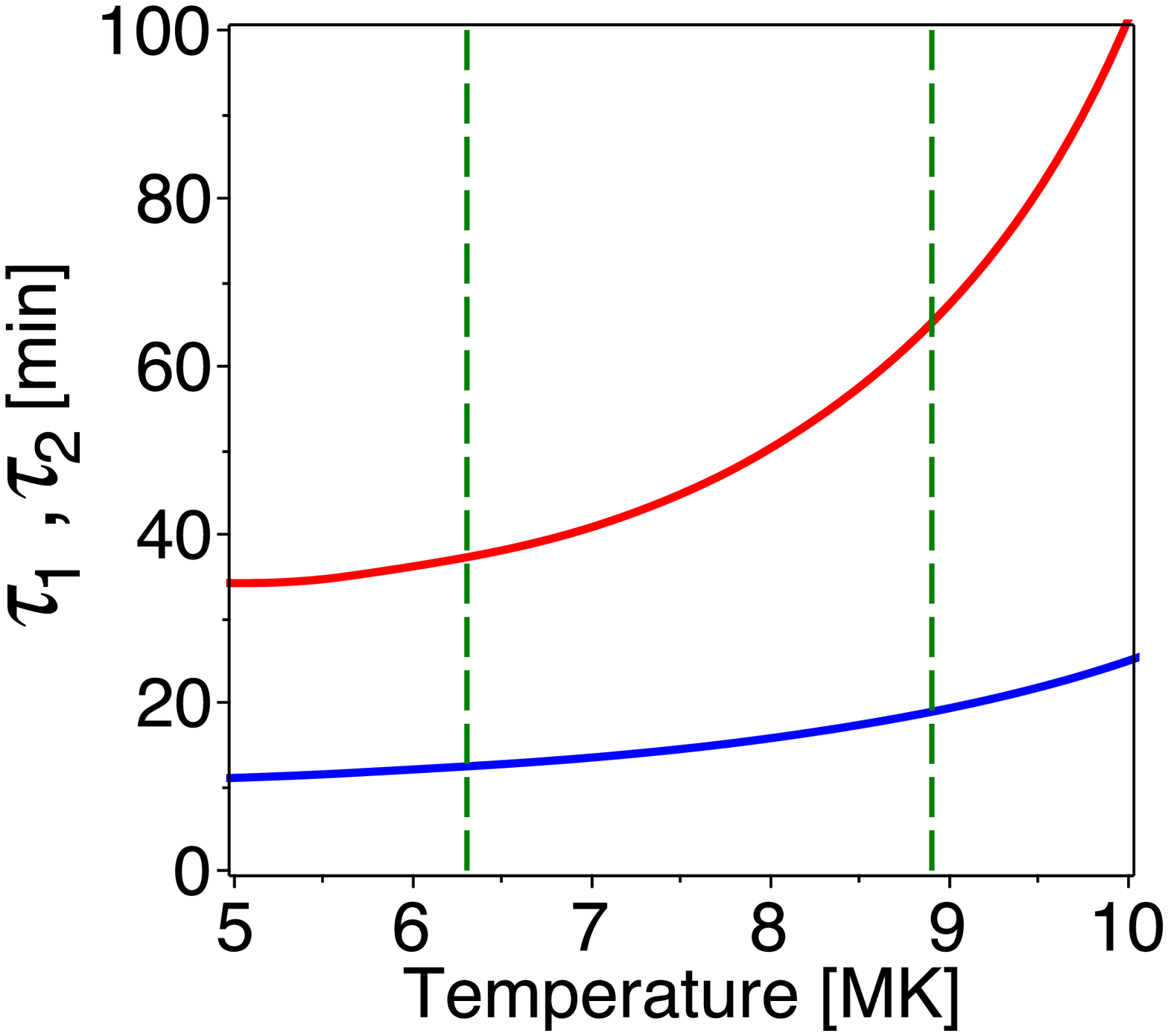}
	\end{center}
	\caption{Left: {A piecewise} dependence of the optically thin radiation losses {per unit mass $L(\rho,T)=\chi \rho T^\beta$} on temperature, {where the specific values of the parameters $\chi$ and $\beta$ are} determined from the CHIANTI atomic database v. 8.0.7 for the plasma concentration $10^{16}$\,m$^{-3}$, {and vary with the temperature interval considered.}
		Right: Variation of $\tau_1$ (red) and $\tau_2$ (blue) determined by Eq.~(\ref{eq:tau1_2}) with temperature, for the radiative cooling shown in the left-hand panel and some heating model with the density and temperature power indices $a=-0.5$ and $b=-3$, respectively. The green dashed lines indicate the SUMER observational channels 6.3\,MK and 8.9\,MK.
	}
	\label{fig:chianti}
\end{figure}

For the solar corona, the optically thin radiation loss function can be modelled as $L(\rho,T)=\chi \rho T^\beta$, whose temperature dependence is illustrated in Fig.~\ref{fig:chianti}, determined from the CHIANTI atomic database \citep{1997A&AS..125..149D, 2015A&A...582A..56D}.
{Function $L(\rho,T)$ represents the radiative losses per unit mass (W kg$^{-1}$), which is obtained from the radiative losses per unit volume (W m$^{-3}$) divided by the plasma density $\rho$}.
Likewise, the unknown coronal heating function can be locally parametrised as $H(\rho,T)=h \rho^a T^b$ {\citep[see e.g.][]{1978ApJ...220..643R, 1993ApJ...415..335I,1988SoPh..117...51D}}, where a certain combination of the power law indices $a$ and $b$ could be associated with a specific heating mechanism. The proportionality coefficient $h$ can in turn be determined applying the thermal equilibrium condition $Q(\rho_0,T_0) = 0$.
{More recent observational and theoretical works suggested that the coronal heating function may also have an intermittent time-dependent component \citep[see e.g.][]{2006SoPh..234...41K,2016ApJ...826L..20R}. Characteristic times of such a time-varying heating are shown to be predominantly short, shorter than a minute \citep[e.g.][]{2014Sci...346B.315T,2016ApJ...816...12T}. On the time scales of the considered slow coronal waves (with periods from several minutes to several tens of minutes), the chosen form of the function $H(\rho,T)$ thus represents a time-averaged steady heating, sustaining the oscillating loop at approximately the same mean temperature.}
Thus, we determine a misbalance between the heating and cooling processes in the solar corona, caused by slow waves, through different dependences of the functions $L(\rho,T)$ and $H(\rho,T)$ on the plasma density and temperature perturbed by the wave.
{As a specific heating {scenario} has not been revealed yet, the power law indices $a$ and $b$ in the parametric dependence of the heating function are treated as free parameters.}

We linearise the governing equations around the initial equilibrium, obtaining energy equation (\ref{eq:energy_full}) in the form
\begin{equation}\label{eq:energy_lin}
\frac{\partial \tilde{T}}{\partial t} - (\gamma-1)\frac{T_0}{\rho_0}\frac{\partial \tilde{\rho}}{\partial t}=\frac{\kappa}{\rho_0 C_\mathrm{V}}\frac{\partial^2\tilde{T}}{\partial z^2} - \frac{\tilde{T}}{\tau_2}-\left(\frac{1}{\tau_2}-\frac{\gamma}{\tau_1}\right)\frac{T_0}{\rho_0}\tilde{\rho},
\end{equation}
where {the symbol \lq\lq$\sim$\rq\rq\ indicates the linear perturbations, and}
\begin{equation}\label{eq:tau1_2}
\tau_{1}={\gamma C_\mathrm{V}}/\left[{Q_{T}-(\rho_0/T_0)Q_{\rho}}\right],\\
\tau_{2}={C_\mathrm{V}}/{Q_{T}}
\end{equation}
are characteristic time scales of the thermal misbalance, fully determined by the parameters of the equilibrium and by the rates of change of the heating/cooling function $Q(\rho,T)$ with density, $Q_\mathrm{\rho}\equiv(\partial Q/\partial \rho)_T $, and temperature $Q_{T}\equiv(\partial Q/\partial T)_\rho $.
In the following analysis, we consider only positive values of both $\tau_1$ and $\tau_2$, thus focusing on the effect of the slow wave (isentropic) damping or over-stability \citep[see][for details]{1965ApJ...142..531F}.
Typical values of the misbalance time scales $\tau_1$ and $\tau_2$ for the radiative cooling determined by CHIANTI and a guessed heating function (determined by the specific values of the density and temperature power indices $a$ and $b$) {in a dense loop \citep{2014LRSP...11....4R}} are illustrated in the right-hand panel of Fig.~\ref{fig:chianti}. For example, for the temperatures associated with SUMER oscillations, 6.3\,MK and 8.9\,MK, we obtain $\tau_1\approx37$\,min and $\tau_2\approx12$\,min, and $\tau_1\approx65$\,min and $\tau_2\approx19$\,min, respectively, for $a=-0.5$ and $b=-3$. This example is provided for the illustrative purposes only, while a more comprehensive analysis of the behaviour of $\tau_1$ and $\tau_2$ with $a$ and $b$ and their effect on the slow wave dynamics are given in Sec.~\ref{sec:stab}. No further assumptions on the values of the characteristic times $\tau_1$ and $\tau_2$ are made in the following analysis, implying the non-adiabatic terms on the right-hand side of energy equation (\ref{eq:energy_lin}) are allowed to be arbitrarily large \citep[in contrast with][where the effect of the thermal misbalance on slow waves is investigated under the assumption of a weak non-adiabaticity]{2016ApJ...824....8K,2017ApJ...849...62N}.
{We would also like to stress that in contrast to previous works \citep[e.g.][]{2004A&A...415..705D}, investigating effects of the radiative cooling on the damping of slow waves keeping the heating term constant, i.e. not affected by the perturbations of the plasma parameters by a wave and hence not contributing into the wave dynamics, we account for the variation of both heating and radiative cooling  by the wave. Therefore, the heating/cooling misbalance times $\tau_{1,2}$ (\ref{eq:tau1_2}) are not associated with the corresponding time scales of the cooling or heating processes considered separately of each other.}

We seek a solution of the linearised set of governing equations in the form $e^{i(kz-\omega t)}$, which yields the following dispersion relation between the cyclic frequency $\omega$ and the wavenumber $k$,
\begin{equation}\label{eq:dr}
\omega^3 + A(k)\omega^2 + B(k)\omega + C(k) = 0,
\end{equation}
where the coefficients are
\begin{align*}
&A=i\left[\frac{k^2\kappa}{\rho_0 C_\mathrm{V}}+\frac{1}{\tau_2}\right],
B=-C_\mathrm{s}^2k^2,C=-i\frac{k_\mathrm{B}T_0}{m}k^2 \left[\frac{k^2\kappa}{\rho_0 C_\mathrm{V}}+\frac{\gamma}{\tau_1}\right],
\end{align*}
where $C_\mathrm{s}=\sqrt{\gamma k_\mathrm{B}T_0/m}$ is a standard definition of the sound speed. 
{We need to mention here that as the plasma gets perturbed by the wave, the condition of the initial isothermality discussed above is violated, allowing the plasma temperature to vary with both space and time. Thus, $C_\mathrm{s}$ is the sound speed in a non-isothermal medium with the adiabatic index $\gamma=5/3$.}
Equation~(\ref{eq:dr}) is found to be asymmetric with respect to space and time, being a fourth- and third-order equation with respect to $k$ and $\omega$, respectively.
Similarly to \citet{2003A&A...408..755D}, a wavelength-dependent term in the coefficients $A(k)$ and $C(k)$ could be associated with the characteristic time scale of the field-aligned thermal conductivity, so that
\begin{equation}\label{eq:taucond}
\tau_{\mathrm{cond}}={\rho_0 C_\mathrm{V}\lambda^2}/{\kappa},
\end{equation}
with $\lambda = 2\pi/k$ being the wavelength.
In the regime of a weak non-adiabaticity, i.e. assuming the parameters $1/\omega\tau_\mathrm{cond}$ and $1/\omega\tau_{1,2}$ are small, dispersion relation (\ref{eq:dr}) reduces to
\begin{equation}\label{eq:dr_weak}
\omega^2 = C_\mathrm{s}^2 k^2\left\{1-i\omega^{-1}\left[\frac{\gamma - 1}{\gamma}\frac{4\pi^2}{\tau_\mathrm{cond}}+\frac{\tau_1-\tau_2}{\tau_1\tau_2}\right]\right\},
\end{equation}
Weakly non-adiabatic dispersion relation (\ref{eq:dr_weak}) is a limiting case of Eq.~(21) in \citet{2017ApJ...849...62N} in neglecting the effects of the viscosity and oblique propagation. In the following analysis, we study full dispersion relation (\ref{eq:dr}).
{Thus, we allow the imaginary part of the frequency to be of the same order of magnitude as the real part. This regime is motivated by the apparently high damping rates of coronal slow oscillations usually observed {(see Sec.~\ref{sec:stab} for references)}.}

\section{Stability analysis}
\label{sec:stab}

Processes described by dispersion relation similar to (\ref{eq:dr}) have been previously shown to affect both the phase speed and the damping/amplification length of propagating magnetoacoustic waves \citep[][]{1992ApJ...396..717I,1993ApJ...415..335I}. In this section we analyse these effects on standing slow magnetoacoustic waves in hot coronal loops (SUMER oscillations), addressing recent advances in observational detections of these waves \citep{2011SSRv..158..397W}. In particular, SUMER oscillations are usually seen to rapidly damp, with the quality factor (\emph{q-factor}) that is the ratio of the damping time to the oscillation period, being less than 2--3 \citep{2003A&A...406.1105W,2006ApJ...639..484M,2016ApJ...830..110C, 2019ApJ...874L...1N}.

Dictated by the observational properties of standing slow oscillations in the corona, we choose the following set of physical parameters
\begin{equation}\label{eq:pars}
\begin{cases}
T_0= 6.3\times 10^6\,\mathrm{K},\\
\rho_0 = 10^{-11}\,\mathrm{kg\,m}^{-3},\\
L=180\times 10^6\,\mathrm{m},\\
\kappa=10^{-11}T_0^{5/2}\, \mathrm{W\,m}^{-1}\,\mathrm{K}^{-1},\\
m=0.6\times1.67\times10^{-27}\,\mathrm{kg},\\
k_\mathrm{B}=1.38\times10^{-23}\,\mathrm{m}^2\,\mathrm{kg}\,\mathrm{s}^{-1}\,\mathrm{K}^{-1},\\
\gamma =5/3,
\end{cases}
\end{equation}
where $L$ is the loop length, and the chosen value of the temperature $T_0$ corresponds to a typical detection of a SUMER oscillation {\citep[see][for the most recent review]{2019ApJ...874L...1N}}. The set of parameters (\ref{eq:pars}) {corresponds to the observations of dense loops \citep[e.g.][]{2017A&A...600A..37N},}
providing the sound speed $C_\mathrm{s}\approx152\sqrt{T_0[\mathrm{MK}]}\approx 382$\,km\,s$^{-1}$, acoustic oscillation period $P=2L/C_\mathrm{s}\approx 15.7$\,min, and the characteristic time scale of the thermal conduction $\tau_\mathrm{cond}\approx448$\,min {(obtained by substitution of the set of parameters (\ref{eq:pars}) into Eq.~(\ref{eq:taucond}) and taking $\lambda = 2L$)}. The ratio of the oscillation period to thermal conduction time, $P/\tau_\mathrm{cond} \approx 0.035$, coincides by an order of magnitude with the estimation in e.g. \citet{2003A&A...408..755D} for the chosen value of $\rho_0$.
{Such a ratio of the oscillation period to the thermal conduction time justifies a non-isothermal nature of the discussed waves, implying that in the considered physical conditions (\ref{eq:pars}) the thermal conduction mechanism is insufficient to smooth out the temperature perturbation on the wave period. However, in shorter and hotter loops the thermal conduction time could be significantly shorter, making the waves almost isothermal.}
In turn, the heating/cooling times $\tau_{1,2}$ (\ref{eq:tau1_2}) are treated as free parameters in this analysis, being mainly determined by the properties of an unknown heating function.

\begin{figure}
	\begin{center}
		\includegraphics[width=0.49\linewidth]{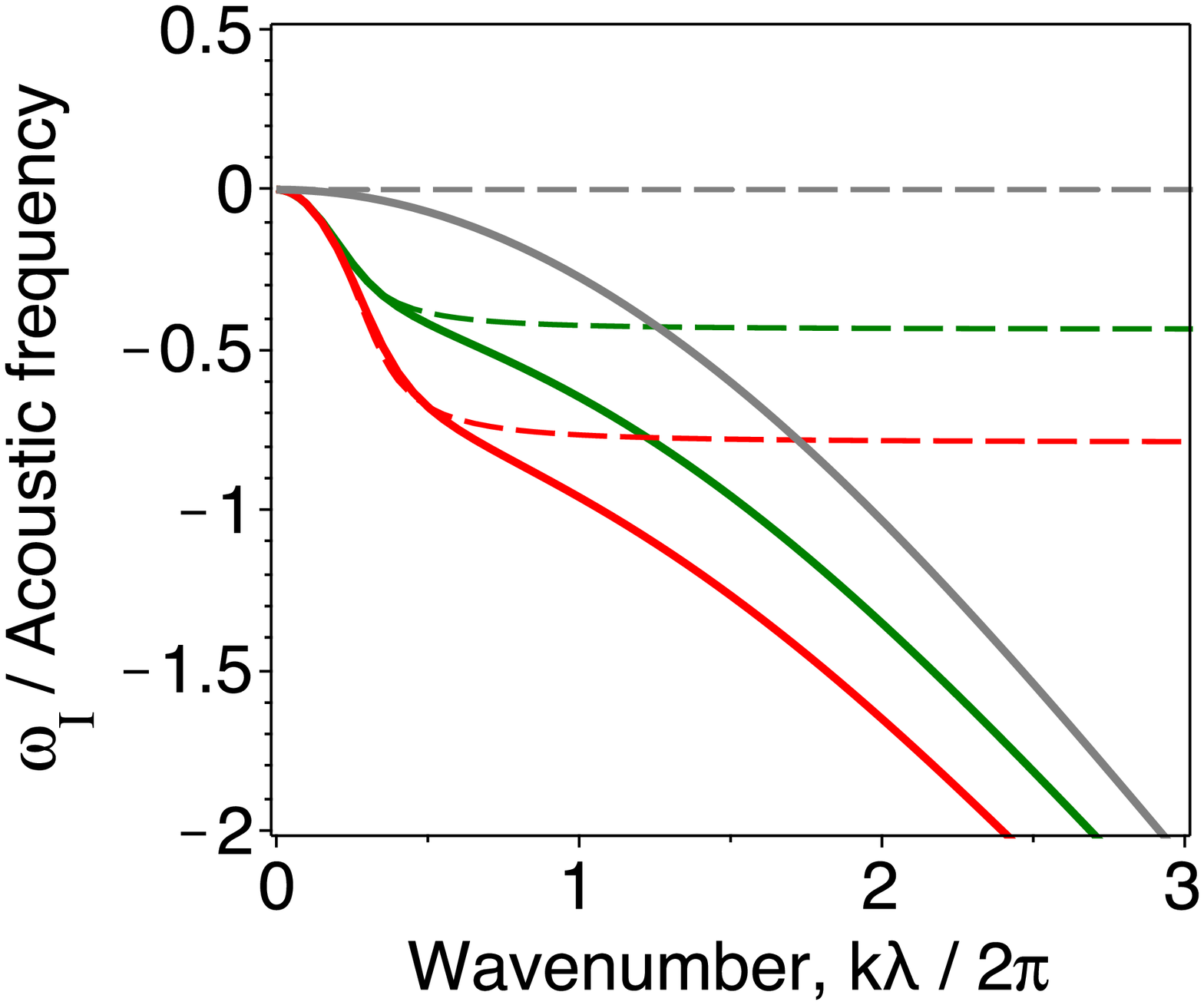}
		\includegraphics[width=0.49\linewidth]{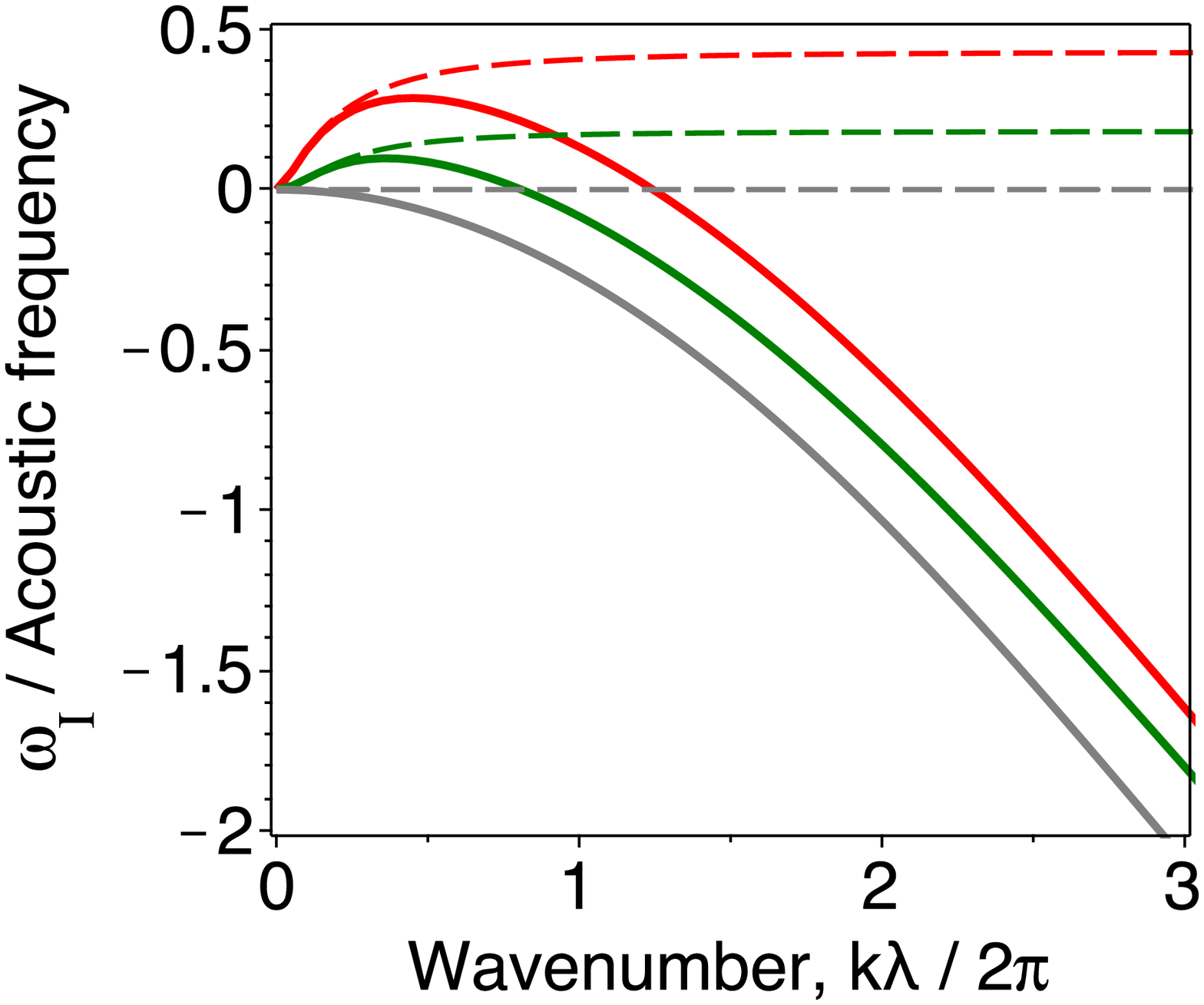}
	\end{center}
	\caption{Variation of $\omega_\mathrm{I}$ obtained for $\tau_1=15$\,min, and $\tau_2=8.2$\,min (green) and $\tau_2=6$\,min (red), left; and $\tau_1=10$\,min, and $\tau_2=13$\,min (green) and $\tau_2=22$\,min (red), right. The grey lines in both panels indicate $\omega_\mathrm{I}^\mathrm{cond}$ obtained with $\tau_{1,2} \to \infty$. The dashed lines in both panels indicate $\omega_\mathrm{I}^\mathrm{M}$ obtained with $\tau_\mathrm{cond} \to \infty$.
	}
	\label{fig:omega}
\end{figure}

We seek a solution to dispersion relation (\ref{eq:dr}) in a standing wave form, i.e. assuming the cyclic frequency $\omega$ to be complex, $\omega=\omega_\mathrm{R}+i\omega_\mathrm{I}$, while the wavenumber $k$ is real. Substituting this into Eq.~(\ref{eq:dr}), we solve the polynomial equation for $\omega_\mathrm{I}$ numerically using \emph{Maple 2016}{\footnote{\url{https://www.maplesoft.com/support/help/}}} environment. Variation of $\omega_\mathrm{I}$ with $k$ is shown in Fig.~\ref{fig:omega} for different values of the heating/cooling times $\tau_{1,2}$, including the case with $\tau_{1,2} \to \infty$ which corresponds to the damping by thermal conduction only, $\omega_\mathrm{I}^\mathrm{cond}$, and with $\tau_\mathrm{cond} \to \infty$ indicating a pure thermal misbalance case, $\omega_\mathrm{I}^\mathrm{M}$. Depending on the values of $\tau_{1,2}$, the imaginary value $\omega_\mathrm{I}^\mathrm{M}$ can contribute either positively or negatively into $\omega_\mathrm{I}^\mathrm{cond}$, revealing regimes of the enhanced damping ($\omega_\mathrm{I}<\omega_\mathrm{I}^\mathrm{cond}$) or suppressed damping ($\omega_\mathrm{I}^\mathrm{cond}<\omega_\mathrm{I}<0$) and over-stability ($\omega_\mathrm{I}>0$). These regimes have been discussed in, e.g. \citet{2016ApJ...824....8K} and \citet{2017ApJ...849...62N}. However, in those works the non-adiabatic effects were weak, thus not describing the strong damping detected in observations {\citep[e.g.][]{2003A&A...406.1105W,2006ApJ...639..484M,2016ApJ...830..110C, 2019ApJ...874L...1N}}.

\begin{figure}
	\begin{center}
		\includegraphics[width=0.48\linewidth]{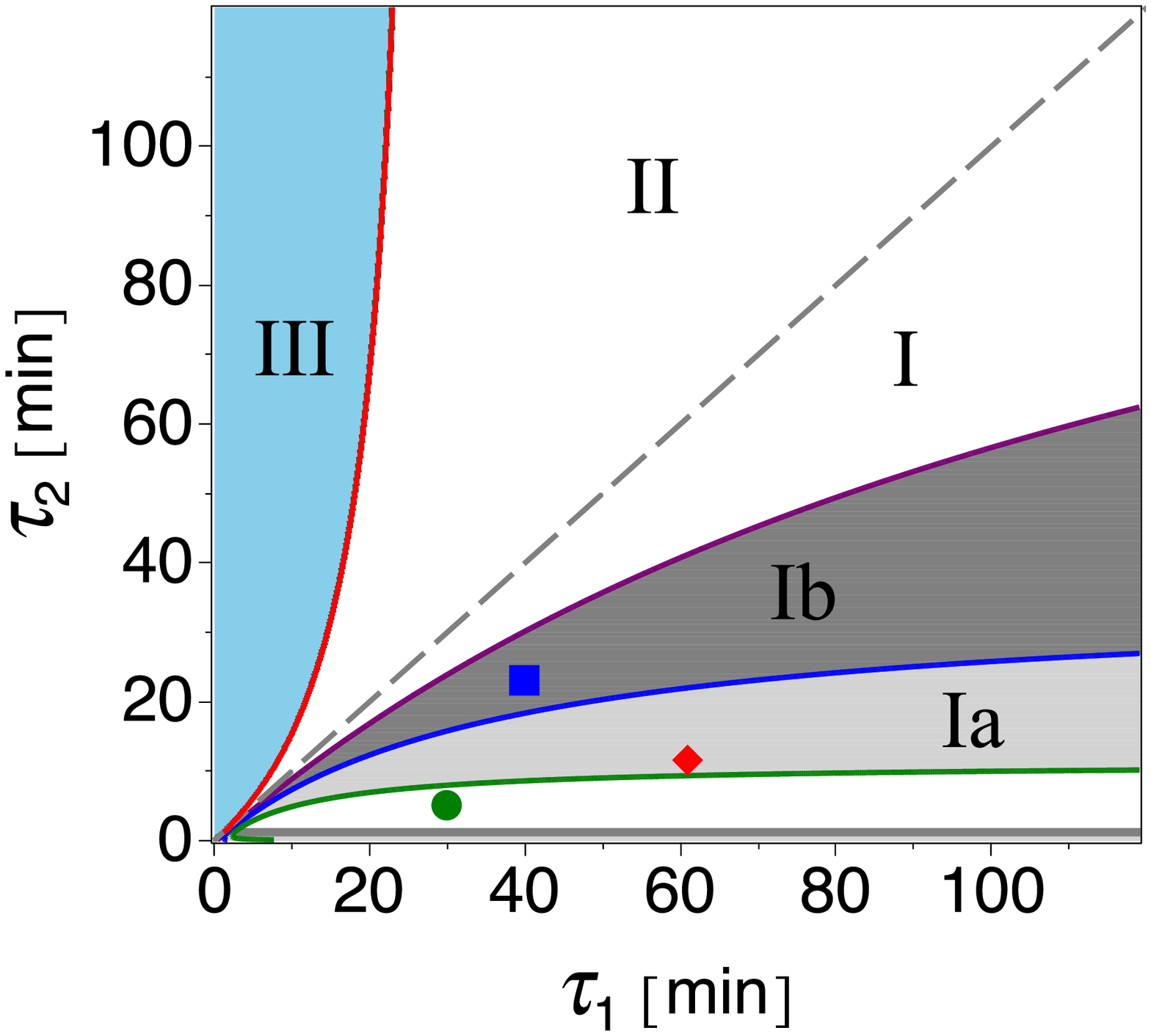}
		\includegraphics[width=0.5\linewidth]{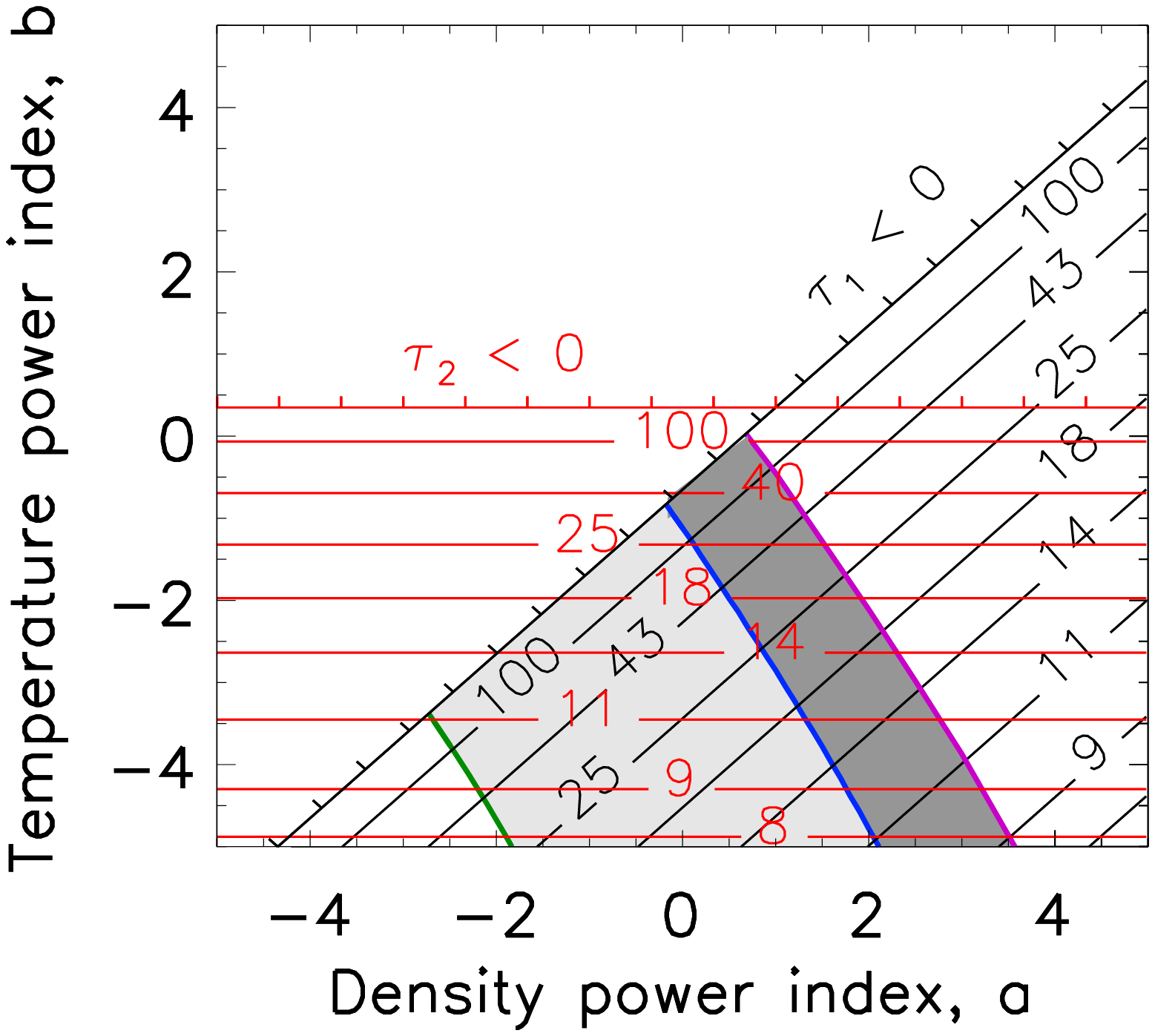}
	\end{center}
	\caption{Left: Parametric regions of the wave damping enhancement (I), suppression (II), and thermal over-stability (III). Grey-shaded regions indicate the values of $\tau_{1,2}$ where the q-factor is in between 1 (the green line) and 2 (the blue line, Ia), and in between 2 and 3 (the purple line, Ib).
		%The orange symbols indicate the values of $\tau_{1,2}$ obtained with the radiative cooling function determined by CHIANTI (see Fig.~\ref{fig:chianti}) and the heating mechanisms given in Table~\ref{tab:heating_models}.
		The red, green, and blue symbols indicate some arbitrary values of $\tau_{1,2}$ chosen for the numerical solutions shown in Fig.~\ref{fig:waves}.
		Right: Heating/cooling times $\tau_{1,2}$ (see Eq.~(\ref{eq:tau1_2}), black and red contours, respectively) determined for the CHIANTI radiative cooling, and the heating function in the form $H(\rho,T) \propto \rho^a T^b$ for the varying temperature and density power indices $a$ and $b$. {The grey-shaded areas indicate the values of $a$ and $b$ where $1 <$\,q-factor\,$<2$ (light grey) and $2 <$\,q-factor\,$<3$ (dark grey). The green, blue, and purple lines show q-factor equals 1, 2, and 3, respectively.}
		% to be from 1 (the green line) to 2 (the blue line).
		%Both panels are obtained with the physical parameters given in (\ref{eq:pars}).
	}
	\label{fig:t1t2}
\end{figure}

The left-hand panel of Fig.~\ref{fig:t1t2} illustrates regions of the damping enhancement, suppression, and over-stability in the two-dimensional parametric space $(\tau_1,\tau_2)$, for the fundamental mode of the oscillation, i.e. with $k=\pi/L$. Here, we treat the characteristic times $\tau_1$ and $\tau_2$ as free parameters. The damping enhancement occurs when $\tau_1>\tau_2$ (see e.g. the last term on the right-hand side of Eq.~(\ref{eq:dr_weak})), where the q-factor drops down to the observational values of about 1--3 {\citep[e.g.][]{2003A&A...406.1105W,2006ApJ...639..484M,2016ApJ...830..110C, 2019ApJ...874L...1N}}. We calculated the values of the heating/cooling times $\tau_{1,2}$ 
%using the CHIANTI model for radiative cooling (see Fig.~\ref{fig:chianti}), and 
adapting four heating models from \citet{1993ApJ...415..335I} (see Table~\ref{tab:heating_models}). For the chosen set of parameters (\ref{eq:pars}), the obtained values of $\tau_{1,2}$ for those heating models are found to be either of different signs or both negative, which would result into the development of thermal instabilities of a non-acoustic nature \citep[see][]{1965ApJ...142..531F}.  Therefore, neither of them is found to be suitable for the observational damping of SUMER oscillations.

\begin{table}
	\centering
	\caption{Coronal heating functions modelled as $H(\rho,T) \propto \rho^a T^b$: Ohmic heating (1), constant heating per unit volume (2) and mass (3), and by Alfv\'en waves/mode conversion (4) \citep[see][]{1993ApJ...415..335I}; and the corresponding $\tau_{1,2}$ (\ref{eq:tau1_2}) in minutes with the radiative cooling determined by CHIANTI (see Fig.~\ref{fig:chianti}).
	}
	\label{tab:heating_models}
	\small\addtolength{\tabcolsep}{-2.7pt}
	\begin{tabular}{ccccc ccccc}
		\hline
		\textbf{Model} & $a$ & $b$ & $\tau_1$&$\tau_2$& \textbf{Model} & $a$ & $b$ & $\tau_1$&$\tau_2$\\
		\hline
		\textbf{1}&0&1&$-42.3$&$-64.6$&\textbf{3}&0&0&$-107.6$&118.9\\
		\textbf{2}&-1&0&$-42.3$&118.9&\textbf{4}&1/6&7/6&$-42.3$&$-51.4$\\
		%3&0&0&$-107.6$&118.9\\
		%4&1/6&7/6&$-42.3$&$-51.4$\\
		\hline
	\end{tabular}
\end{table}

As $\tau_{1,2}$ depend on the parameters $a$ and $b$ of the heating function (\ref{eq:tau1_2}), we calculate $\tau_{1,2}$ for $a$ and $b$ both ranging from e.g. $-5$ to $5$ (see the right-hand panel of Fig.~\ref{fig:t1t2}). The obtained values of $\tau_{1,2}$ are seen to depend strongly on $a$ and $b$, {varying from several to a hundred of minutes and longer for the chosen values of the plasma density and temperature.} They both have the vertical asymptote at $a\approx 1$ and $b\approx 0.4$, above which they both become negative. The blank regions in the right-hand panel of Fig.~\ref{fig:t1t2} and where the contour lines do not intersect correspond to the negative or different signs of $\tau_{1,2}$, respectively, which give raise to other thermal instabilities \citep[see][]{1965ApJ...142..531F} which are out of the scope of this study. 
%The latter case means the temperature and density derivatives of the heating function tend to zero, and the thermal misbalance times are mainly determined by the radiative loss function.
We now compare this diagram to the values of $\tau_{1,2}$, for which the oscillation q-factor was found to vary from 1 to 2 (see the left-hand panel of Fig.~\ref{fig:t1t2}), constraining the heating functions which are able to reproduce the observational damping (see the grey-shaded area in the right-hand panel of Fig.~\ref{fig:t1t2}).
{For lower plasma densities, the values of indices $a$ and $b$, which give the misbalance times $\tau_{1,2}$ about the observed periods, would be even lower.}
%, thus suggesting a new potential way for seismological diagnostics of the parameters of the coronal heating function through the damping of standing slow oscillations.

\begin{figure}
	\begin{center}
		\includegraphics[width=0.49\linewidth]{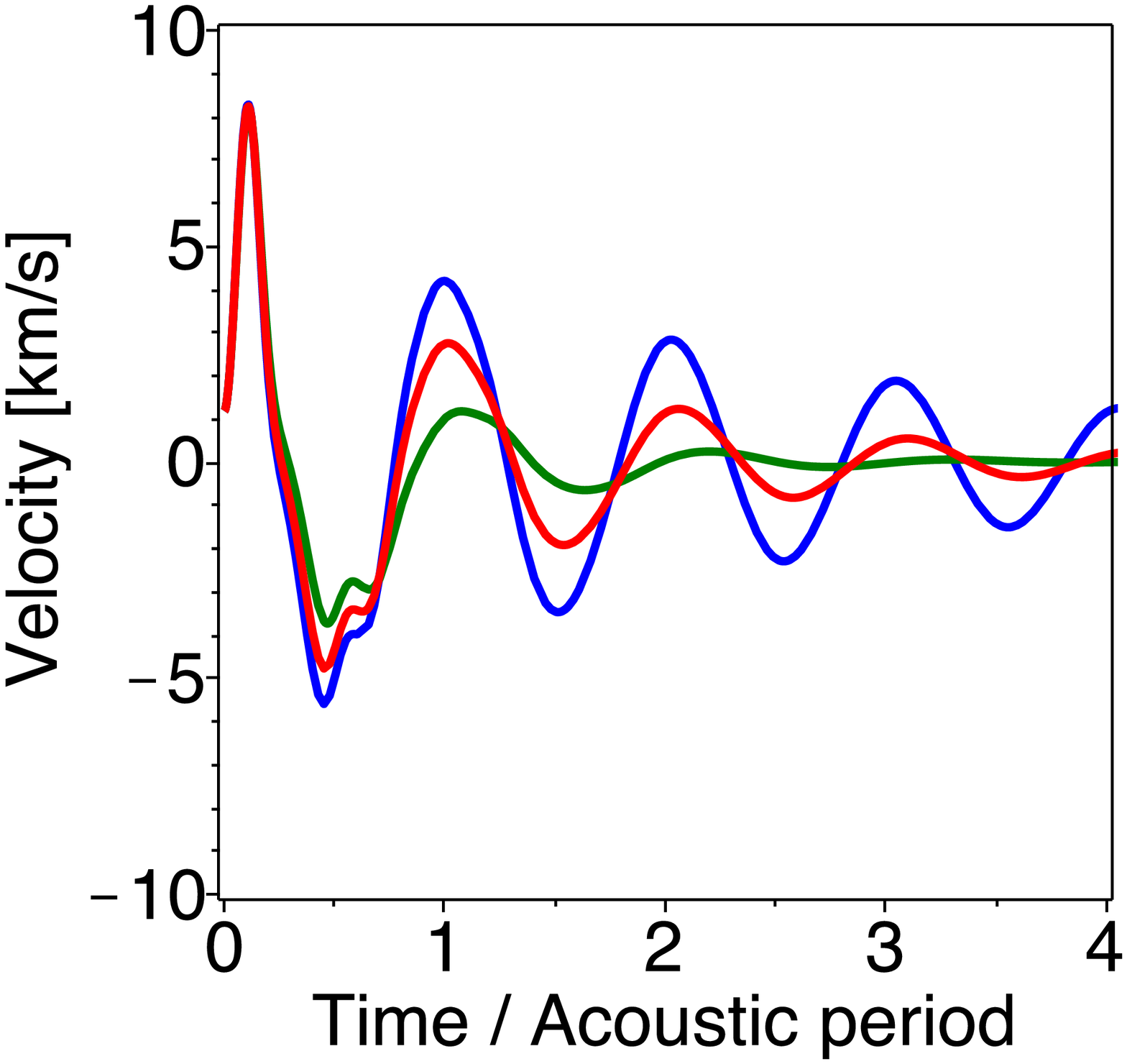}
		\includegraphics[width=0.49\linewidth]{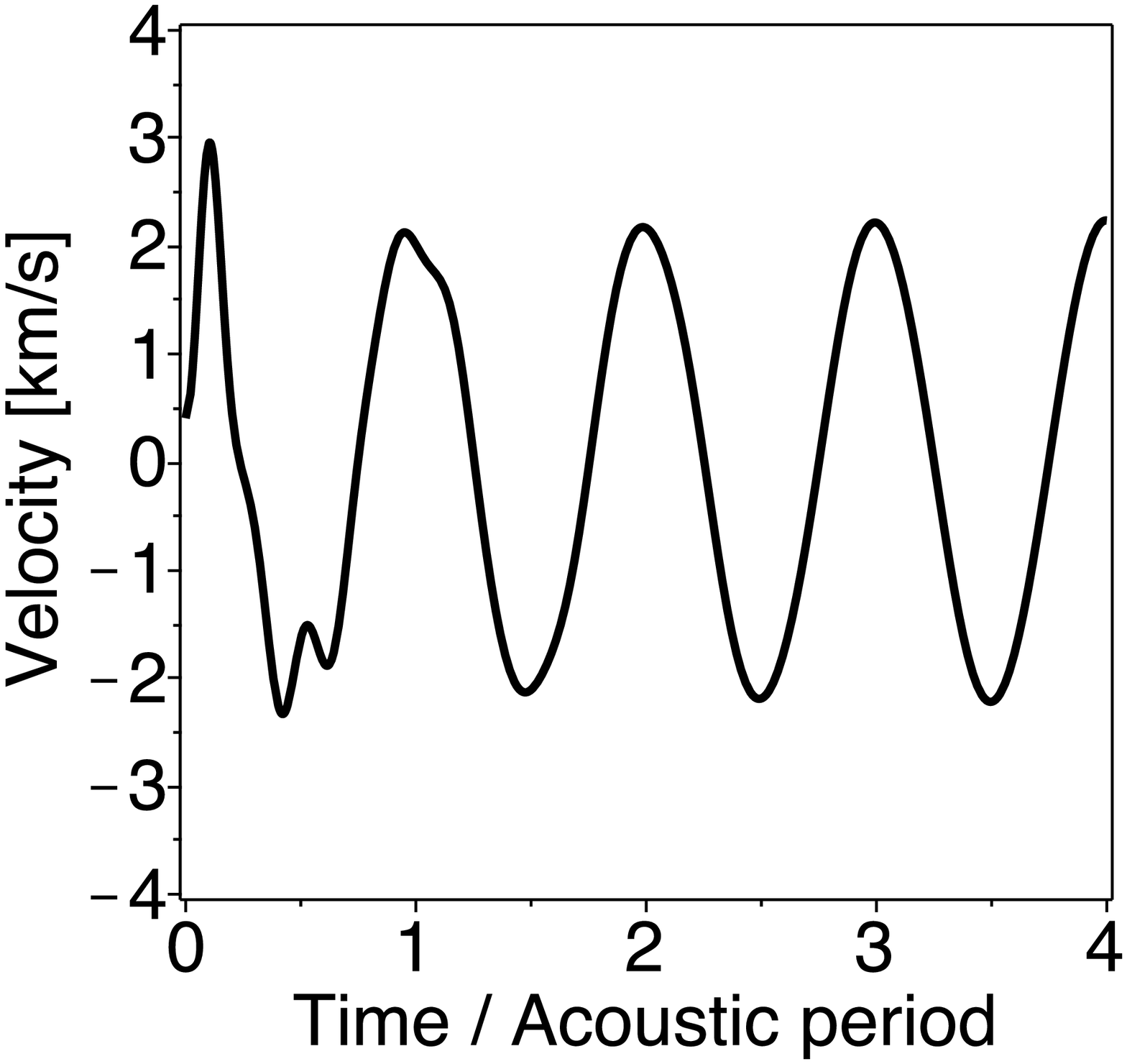}
	\end{center}
	\caption{Left: Cross-sections of the perturbed velocity as a function of time, obtained for $\tau_1 = 40$\,min and $\tau_2 = 23$\,min (blue), $\tau_1 = 61$\,min and $\tau_2 = 11.5$\,min (red), and $\tau_1 = 30$\,min and $\tau_2 = 5$\,min (green).
		%The chosen pairs of the heating/cooling times $\tau_{1,2}$ provide the oscillation q-factor to be from 2 to 3, from 1 to 2, and lower than 1, respectively (see Fig.~\ref{fig:t1t2}, the left-hand panel).
		%The chosen values of $\tau_{1,2}$ providing $1<$~q-factor~$<2$ correspond to the heating model with the density and temperature power indices $a\approx 3.5$ and $b\approx 4$ (see Fig.~\ref{fig:t1t2}, the right-hand panel).
		Right: Similar to the left-hand panel, but for the set of physical parameters from \citet{2008ApJ...681L..41M}, see Sec.~\ref{sec:stab} for details, and with $\tau_1=19.5$\,min and $\tau_2 = 22.3$\,min.
		% corresponding to the heating model with $a = -1$ and $b = 0$ (constant per unit volume).
	}
	\label{fig:waves}
\end{figure}

Choosing three different pairs of the heating/cooling times $\tau_1$ and $\tau_2$, which provide the q-factor to be lower than 1, from 1 to 2, and from 2 to 3, and using parameters (\ref{eq:pars}), we solve the linearised set of governing equations numerically in \emph{Maple 2016}, in a closed resonator located between $z=0$ and $z=L$ and with the initial broadband Gaussian-shaped acoustic perturbation of the width $w=0.12L$, shifted towards one of the boundaries. The cross-sections of the obtained standing solutions at $z=L/2$ are shown in the left-hand panel of Fig.~\ref{fig:waves}. As expected from the dispersion relation (see Eq.~(\ref{eq:dr}) and Fig.~\ref{fig:omega}), the higher harmonics decay faster, so that after about one cycle of the oscillation the initial broadband pulse develops into a pure fundamental mode which then also decays. This example illustrates how sensitive the damping of standing slow waves is to the parameters of the heating/cooling function, and it represents the rapidly decaying oscillations of the SUMER-oscillation type. In a more exotic case, when the values of $\tau_1$ and $\tau_2$ appear to be just near the boundary $\omega_\mathrm{I}=0$ (see e.g. the red line in the left-hand panel of Fig.~\ref{fig:t1t2}), the damping could be highly suppressed by the thermal misbalance \citep[see e.g. the apparently non-decaying oscillation observed in the \ion{Fe}{xv} emission line in Fig.~3 of][]{2008ApJ...681L..41M}. Adapting the physical parameters corresponding to this observation, namely $T_0=10^{6.32}$\,K and $L=342$\,Mm, and choosing $\rho_0 = 10^{-12}$\,kg\,m$^{-3}$, we can reproduce the observed non-decaying oscillation within the developed model for, e.g. $\tau_1=19.5$\,min and $\tau_2 = 22.3$\,min.
%corresponding to the constant per unit volume heating function with the temperature and density power indices $a = -1$ and $b = 0$.

\section{Summary and conclusions}
\label{sec:sumcon}

We investigated the mechanism for damping of {linear} standing slow magnetoacoustic waves in the solar corona through the misbalance of some heating and radiative cooling processes.
{We addressed the coronal part of a loop with an isothermal equilibrium. This is a standard approach for modelling slow waves in the corona. However, we consider the wave dynamics in the presence of a temperature- and pressure-depend heating and radiative cooling and thermal conduction, addressing a misbalance of those processes caused by the waves.}
The wave dynamics was found to be highly sensitive to the parameters of the misbalance, expressed in terms of the characteristic times $\tau_{1,2}$ of the heating/cooling function change with the plasma density and temperature perturbed by the wave (see Sec.~\ref{sec:gov}). Depending upon the values of $\tau_{1,2}$, we found three different regimes of the wave evolution, which are the enhanced and suppressed damping (with respect to the one caused by the field-aligned thermal conductivity), and the thermal over-stability. Unlike the previous analytical works, we did not treat the non-adiabatic terms small, that allowed us to obtain the enhanced damping rates matching those detected in observations. 

Our findings allow one to reproduce the observed behaviour of SUMER oscillations, keeping the thermal conduction coefficient in its standard estimation, but accounting for the heating/cooling misbalance. For the set of physical parameters corresponding to the observations of SUMER oscillations (see Sec.~\ref{sec:stab}), the characteristic time scale of the thermal conduction was found to be at least an order of magnitude longer than the oscillation period. This indicates a low efficiency of the field-aligned thermal conductivity in damping these oscillations. In turn, typical heating/cooling times $\tau_{1,2}$ were found to be comparable to the observed periods of SUMER oscillations (from a few minutes to a few tens of minutes, see Fig.~\ref{fig:t1t2}), for a sufficiently broad range of the heating function parameters and for the CHIANTI radiative cooling. For $\tau_{1}>\tau_{2}$, this results into a domination of the damping by the heating/cooling misbalance over conductive damping.
{Moreover, the discussed effect persists even in the limiting case of isothermal waves, which are not subject to the damping by thermal conduction at all, occurring in the case of the dominant thermal conduction \citep{2003A&A...408..755D}. In this regime, the cooling and heating functions, and hence their misbalance, are still affected by the perturbations of density in the wave and hence contribute to its damping.}

Using the CHIANTI model for the radiative cooling and fixing other parameters of the equilibrium, the values of $\tau_{1,2}$ become fully determined by the heating function. This suggests a new way for the diagnostics of the coronal heating mechanism via damping of SUMER oscillations. For example, neither of four heating models considered by \citet{1993ApJ...415..335I} (see Table~\ref{tab:heating_models}) was found to reproduce the observed damping of SUMER oscillations. On the other hand, we determined the range of the power-law indices $a$ and $b$, which give the observed damping times.
%The parametric regions for $a$ and $b$ shown in Fig.~\ref{fig:t1t2} thus {may} put additional constraints onto the heating models.
Moreover, the developed theory could also address a more exotic case of an apparently non-decaying SUMER type oscillation detected by \citet{2008ApJ...681L..41M}, by choosing the values of $\tau_1$ and $\tau_2$ which give $\omega_\mathrm{I}\approx0$. In addition, acoustic over-stability could be considered as a mechanism for the excitation of 8--30~min oscillations of the soft X-ray emission generated in pre-flaring active region \citep{2016ApJ...833..206T}.

The need to comply with observational properties of coronal slow waves may put additional constraints on the empirical determination of the dependence of the heating function on the plasma parameters. This seismological information about the acceptable ranges of the parameters $a$ and $b$, together with the information obtained by other methods, could be used for revealing the heating function. In particular, our study suggests that $-2\lesssim a\lesssim 2$ and $b\lesssim 0$ for the chosen values of the equilibrium density and temperature. Those intervals should be subject to a dedicated follow-up analysis. In particular, the effect of different parametric forms of the heating function dependence on the density and temperature, e.g. polynomial, should be considered.
{Likewise, the time-dependence of the coronal heating function, neglected in this study on the time scale of a slow wave, could be more important for shorter-period coronal MHD waves, e.g. the fast waves with about 1-min periodicity.}
{Also, this neglection does not allow us to address the transient events in which the loop is impulsively heated and rapidly cools down at the time scale comparable to the wave period \citep[e.g.][]{2019arXiv190702291R}, thus making the developed theory restricted to the loops sustained at approximately the same mean temperature during the whole wave evolution.}
In addition, the future development of the theory needs to soften certain assumptions made in this paper. In particular, we neglect the effects of the oblique wave propagation{, i.e. the departure of the slow wave speed from the sound speed in the case of finite $\beta$,} and viscosity, which could bring additional time scales into the problem. This could be important if the coronal heating depends on the magnetic field \citep{1992PPCF...34..411H,2017ApJ...849...62N}. We also do not consider the effect of geometrical dispersion \citep{1983SoPh...88..179E,2015ApJ...807...98Y} that is usually weak {for slow waves in coronal loops.} Likewise, we do not account for nonlinear effects. 
{Another interesting development of this study could be the inclusion of a chromosphere.}
Accounting for these effects should be addressed in a follow up study.

\begin{acknowledgements}
This work was supported by STFC consolidated grant ST/P000320/1 (V.M.N., D.Y.K.),  and the Russian Foundation for Basic Research grant No. 18-29-21016  (V.M.N.). Calculations presented in the reported study were also funded by RFBR according to the research project No. 18-32-00344 (D.I.Z.). 
The study was supported in part by the Ministry of Education and Science of Russia under the public contract with educational and research institutions within the project 3.1158.2017/4.6.
CHIANTI is a collaborative project involving George Mason University, the University of Michigan (USA) and the University of Cambridge (UK).
{Maple is a trademark of Waterloo Maple Inc.}
\end{acknowledgements}

\bibliographystyle{aa} % style aa.bst
%\bibliography{aanda_slow_ref} % your references Yourfile.bib

\end{document}